\documentclass[twocolumn]{revtex4-1}
\usepackage[english]{babel}
\usepackage{amsmath, amssymb}
\usepackage{graphicx}
\usepackage{dcolumn}
\usepackage{subfigure}
\usepackage{epstopdf}
\newcommand{\kslash}{\hbox{$k\!\!\!{\slash}$}}
\newcommand{\pslash}{\hbox{$p\!\!\!{\slash}$}}

\begin{document}

\title{\bf One-loop conformal anomaly in an implicit momentum space regularization framework}

\author{A. R. Vieira $^a$}\email[]{arvieira@fisica.ufmg.br}
\author{J. C. C. Felipe $^{b}$}\email[]{guaxu@fisica.ufmg.br}
\author{G. Gazzola $^a$}\email[]{adriano@fisica.ufmg.br}
\author{Marcos Sampaio $^{a, c}$}\email[]{msampaio@fisica.ufmg.br}

\affiliation{$^a$ Universidade Federal de Minas Gerais - Departamento de F\'{\i}sica - ICEX \\ P.O. BOX 702,
30.161-970, Belo Horizonte MG - Brazil}
\affiliation{$^b$ Universidade Federal de Lavras - Departamento de F\'{\i}sica \\ P.O. BOX 3037, 37.200-000, Lavras MG - Brazil}
\affiliation{$^c$ Centre for Particle Theory - Dept of Mathematical Sciences - Durham University\\
Science Laboratories South Rd. Durham DH1 3LE - UK}

\begin{abstract}
In this paper we consider matter fields in a gravitational background in order to compute the breaking of
the conformal current at one-loop order. Standard perturbative calculations of conformal symmetry breaking expressed by the non-zero 
trace of the energy-momentum tensor have shown that some violating terms are regularization dependent, which may suggest the existence 
of spurious breaking terms in the anomaly. Therefore, we perform the calculation in a momentum space regularization framework in which 
regularization dependent terms are judiciously parametrized. We compare our results with those obtained in the literature
and conclude that there is an unavoidable arbitrariness in the anomalous term $\Box R$. 
\pacs{04.62.+v, 11.10.Gh, 11.30.-j}
\end{abstract}

\maketitle

\section{Introduction}

Conformal invariance (CI) imposes strong constraints on correlation functions leading to exact results mainly in two 
dimensions. On the other hand most renormalizable theories possessing  conformal invariance
at the classical level exhibit the trace anomaly once quantum corrections are taken into
account. Of course this is most welcome in particle physics because conformal symmetry breaking must come into play to properly describe 
the real world. Then, in the high energy limit, CI may be recovered appearing as UV and/or IR limits \cite{Livro}. 
Furthermore, CI is an important concept in holographic theories based on the $AdS_{n+1}/CFT_{n}$ duality which relates 
strongly-coupled four dimensional gauge theory to gravitational theory in five dimensional $AdS$ space-time, for instance. It is also 
important in super-symmetric gauge theories, e.g. (conformal invariant) ${\cal{N}}=4$ super Yang-Mills. For applications of the $AdS/CFT$
conjecture \cite{Maldacena} in many branches of physics see \cite{Makoto}.


Anomalies occur when a symmetry presented at a classical level is broken upon quantization. In perturbation theory, during the process of 
regularization/renormalization, counter-terms are generated and may violate the symmetry that was present at the 
classical level. The presence of anomaly depends on the fact that it is not possible to find a regulator that preserves all the symmetries
of the classical action. Well-known examples of anomalies are the (AVV) chiral anomaly \cite{ABJ} when
gauge fields coupled to conserved currents give rise to non-conserved axial current, and the trace
anomaly of a scalar field conformally coupled to a classical gravitational background \cite{Duff1,Brown}.

Finite and undetermined local terms appear as differences between loop integrals with the same degree of divergence in Feynman diagram calculations \cite
{Jackiw}. Such indeterminacies are regularization dependent and are at the heart of symmetry breakings by regularizations.  A reasonable strategy would be to 
leave them arbitrary till the end of the calculation to be fixed on symmetry or physical grounds. Anomalies, such as the AVV chiral anomaly, appear in this 
approach when the ambiguities proved themselves insufficient to preserve the full set of symmetry identities valid at classical level.

Attributing spurious values to such indeterminacies can break gauge invariance or super-symmetry \cite{IRA}. In the latter reference it was shown that undetermined local terms can be cast as surface terms at any loop order. Moreover it was argued that Momentum Routing Invariance (MRI) is a necessary and sufficient condition to preserve (abelian) gauge symmetry at arbitrary loop order. This condition is automatically fulfilled by dimensional regularization \cite{tHooft}. The strategy of identifying ambiguous regularization dependent surface terms in perturbation theory to arbitrary loop order is better understood and accomplished within Implicit Regularization (IR) \cite{IR}, which is discussed in more detail in Section \ref{2}.

IR is a momentum space setting to perform Feynman diagram calculations in a regularization independent fashion. Consequently IR turns out particularly adequate to
unravel anomalies within  perturbation theory. In IR, the  Lagrangian of the underlying quantum field theory is not modified because neither an explicit
regulator is introduced nor the dimensionality of the space time needs to be moved away from its physical dimension. In particular, IR allows for a democratic display of the anomaly 
between the Ward identities which ultimately should be fixed on physical grounds. For example, in \cite{Leo} was studied Weyl fermions on a classical gravitational background in two 
dimensions and shown that, assuming Lorentz symmetry, the Weyl and Einstein Ward identities reduce to a set of algebraic equations for the arbitrary parameters
which place the Ward identities on equal footing, just as in the AVV triangle anomaly \cite{IRA}. 


In this contribution we revisit an old controversy related to breaking of
the conformal current at one-loop order when matter fields lie on a gravitational background. Some of the terms of this anomaly 
are ambiguous and regularization dependent \cite{review,Shapiro2}. Therefore, we investigate if this ambiguity appears as surface terms which sometimes
may be fixed on symmetry grounds. Moreover, we believe that performing this calculation in four dimensions and without introducing
an explicit regulator is worthwhile since we will not get spurious terms that may contaminate the anomaly.

This work is organized as follows: in section \ref{first} we review some aspects about conformal anomaly; in section \ref{2} we outline the implicit regularization scheme to establish 
our notation; in section \ref{3} we derive the $a'$ coefficient using the one-loop correction to the graviton propagator; in section \ref{4} we perform the one-loop  renormalization of 
the quantum effective action; we present how the anomaly is affected by the surface terms in section \ref{s6} and we draw concluding remarks in section \ref{5}. 

\section{Aper\c{c}u on conformal anomaly}\label{first}

In order to present the state of the art let us establish some notation. A theory is  conformal invariant if it does not change under the field transformation
\begin{equation}
\Psi'(x)= e^{d\sigma(x)}\Psi(x),
\label{1.1}
\end{equation}
where $\Psi$ stands for scalar, vector, spinor or the metric ($ \Psi= \phi, A_{\mu}, \psi$ or $g_{\mu\nu}$, respectively), $\sigma$ is an arbitrary scalar field 
and $d$ is the corresponding conformal weight for the scalar, vector or spinor fields($d=-1, 0$ and $-\frac{3}{2}$, respectively) and it is equal 2 for the metric.

The corresponding conserved current associated with the transformation (\ref{1.1}) is the conformal current also known as the trace of the 
energy-momentum tensor. In classical field theory, this current is conserved in the massless limit. Quantum corrections usually  break conformal invariance in 
the semi-classical approach of gravity (see \cite{review} for a review). Pioneering works about this anomaly have derived one-loop corrections to 
the graviton propagator due to vector \cite{Duff1} and spinor \cite{Duff2} couplings. They found out that, although diffeomorphism was 
preserved, the trace of the energy-momentum tensor was no longer zero at the quantum level \cite{Duff3} since it received finite corrections. Like 
other anomalies this breaking poses a renormalizability issue \cite{Duff4}. 

At first, this symmetry breaking was thought as being spurious \cite{Kallosh} -\cite{Adler}, that is to say an artifact of the regularization method, motivating 
the seek for a regularization scheme which preserves both CI and diffeomorphism \cite{Englert}-\cite{Antoniadis3}. Afterwards, the trace of energy-momentum 
tensor was computed in several frameworks. In \cite{Duff3} it was calculated diagrammatically using dimensional regularization \cite{tHooft,Bollini}. Moreover, 
it was shown that the anomaly also arises in $\zeta$-function regularization \cite{Hawking}, 
point-splitting regularization \cite{Christensen} and in the context of Schwinger-DeWitt method \cite{Vilkovisky2}. A derivation based on the AdS$\slash$CFT 
correspondence can be found in \cite{Henningson}. Besides, this anomaly has already been classified in a regularization independent way using the algebraic 
approach\cite{Nicolas}. However, the explicit diagrammatic computation reveals that some of the terms which quantum mechanically break conformal invariance are 
regularization dependent. For a review about conformal anomaly and its universalities and ambiguities in different regularization schemes see \cite{Shapiro2}.

It is noteworthy that the anomalous trace of the energy-momentum tensor has physical consequences: it determines the energy-momentum tensor for a black hole in 
two dimensions \cite{Christensen2} and the classification of the vacuum quantum states in four dimensions \cite{Balsan}. This anomaly also gives rise to the 
stability condition in the modified Starobinsky inflationary model \cite{Starobinsky,Shapiro}. Besides, the anomaly induced action has applications in black hole 
evaporation \cite{Sergei}, annihilation of an AdS universe \cite{Sergei2} and creation of a de Sitter wall universe \cite{Sergei3}.

The trace anomaly has a general form given by
\begin{equation}
T=\left\langle T^{\mu}_{\mu}\right\rangle= a C^2 + c E+ a' \Box R,
\label{1.2}
\end{equation}
where $C^2=R^2_{\mu\nu\alpha\beta}-2R^2_{\alpha\beta}+\frac{1}{3}R^2$ is the square of the Weyl tensor, 
$E=R^2_{\mu\nu\alpha\beta}-4R^2_{\alpha\beta}+R^2$ is the Gauss-Bonnet topological invariant, $R$ is the Ricci scalar and $a,c$ and $a'$ 
are related with $\beta$-functions \cite{Shapiro2}
\begin{eqnarray}
\beta_1=\frac{1}{(4\pi)^2}\left(\frac{1}{120}N_s+\frac{1}{20}N_f+\frac{1}{10}N_v\right),\nonumber\\
\beta_2=\frac{1}{(4\pi)^2}\left(\frac{1}{360}N_s+\frac{11}{360}N_f+\frac{31}{180}N_v\right),\nonumber\\
\beta_3=\frac{1}{(4\pi)^2}\left(\frac{1}{180}N_s+\frac{1}{30}N_f-\frac{1}{10}N_v\right).
\label{1.3}
\end{eqnarray}

The usual results in the literature are $a=\beta_1$ and $c=\beta_2$. However, there is a disagreement in the coefficient $a'$. While some regularization schemes 
predict $a'=\beta_3$, dimensional regularization yields $a'=\frac{2}{3}\beta_1$\cite{Duff3}. Furthermore, $a'$ vanishes in the derivation based on the AdS$\slash$
CFT \cite{Henningson} correspondence and it is ambiguous in Pauli-Villars regularization \cite{Shapiro2,Shapiro5,Shapiro3}.

Afterwards, it was shown that dimensional regularization actually also furnishes an ambiguous result \cite{Shapiro2}.

We shall compute the trace anomaly in an implicit momentum space regularization framework, paying particular attention to regularization 
dependent quantities \cite{Jackiw,IRA}. We perform the 
one-loop correction to the graviton propagator due to couplings with scalar, fermion and vector fields. We then relate that correction 
for the two-point function with $\left\langle T^{\mu}_{\mu}\right\rangle$. For this purpose we employ implicit regularization \cite{IR} 
in which divergences are expressed order by order in perturbation theory as loop integrals in consonance with BPHZ theorem \cite
{Adriano} whereas undetermined regularization dependent local terms are expressed by surface terms. Thus we derive the $a'$ coefficient 
and then compare our result with those of the literature.

\section{Implicit Regularization}
\label{2}

We apply the implicit regularization framework \cite{IR} to treat the integrals which appear in the amplitudes of section \ref{3}. Let 
us make a brief review of the method. In this scheme, we assume the existence of an implicit regulator $\Lambda$ just to justify 
algebraic operations within the integrands. We then use the following identity to separate  UV divergent basic integrals from the finite part:

\begin{align}
\int_k \frac{1}{(k+p)^2-m^2}&=\int_k\frac{1}{k^2-m^2}\nonumber\\
&-\int_k\frac{(p^2+2p\cdot k)}{(k^2-m^2)[(k+p)^2-m^2]},
\label{2.1}
\end{align}
where $\int_k\equiv\int^\Lambda\frac{d^4 k}{(2\pi)^4}$, to separate basic divergent integrals (BDI's) from the finite part. These BDI's are defined as follows
\begin{equation}
I^{\mu_1 \cdots \mu_{2n}}_{log}(m^2)\equiv \int_k \frac{k^{\mu_1}\cdots k^{\mu_{2n}}}{(k^2-m^2)^{2+n}},
\end{equation}
\begin{equation}
I^{\mu_1 \cdots \mu_{2n}}_{quad}(m^2)\equiv \int_k \frac{k^{\mu_1}\cdots k^{\mu_{2n}}}{(k^2-m^2)^{1+n}}
\end{equation}
and
\begin{equation}
I^{\mu_1 \cdots \mu_{2n}}_{quart}(m^2)\equiv \int_k \frac{k^{\mu_1}\cdots k^{\mu_{2n}}}{(k^2-m^2)^{n}}.
\end{equation}

The basic divergences with Lorentz indices can be combined as differences between integrals with the same superficial degree 
of divergence, according to the equations below, which define surface terms \footnote{The Lorentz indices between brackets stand for 
permutations, i.e. $A^{\{\alpha_1\cdots\alpha_n\}}B^{\{\beta_1\cdots\beta_n\}}=A^{\alpha_1\cdots\alpha_{n}}B^{\beta_1\cdots\beta_n}$ 
+ sum over permutations between the two sets of indices $\alpha_1\cdots\alpha_{n}$ and $\beta_1\cdots\beta_n$}:
\begin{align}
&\Upsilon^{\mu \nu}_{2w}= \eta^{\mu \nu}I_{2w}(m^2)-2(2-w)I^{\mu \nu}_{2w}(m^2) 
\equiv \upsilon_{2w}\eta^{\mu \nu},
\label{dif1}\\
\nonumber\\
&\Xi^{\mu \nu \alpha \beta}_{2w}=  \eta^{\{ \mu \nu} \eta^{ \alpha \beta \}}I_{2w}(m^2)
 -\nonumber\\&-4(3-w)(2-w)I^{\mu \nu \alpha \beta }_{2w}(m^2) \equiv  \xi_{2w}\eta^{\{ \mu \nu} \eta^{ \alpha \beta \}},
\label{dif2}\\
\nonumber\\
&\Sigma^{\mu \nu \alpha \beta \gamma \delta}_{2w} = \eta^{\{\mu \nu} \eta^{ \alpha \beta} \eta^ {\gamma \delta \}}I_{2w}(m^2)-\nonumber\\
&-8(4-w)(3-w)(2-w) I^{\mu \nu \alpha \beta \gamma \delta}_{2w}(m^2)\nonumber\\
& \equiv   \sigma_{2w} \eta^{\{\mu \nu} \eta^{ \alpha \beta} \eta^ {\gamma \delta \}},
 \label{dif3}\\
\nonumber\\
&\Omega^{\mu \nu \alpha \beta \gamma \delta \epsilon \zeta}_{2w} =  \eta^{\{\mu \nu} \eta^{ \alpha \beta} \eta^ {\gamma \delta}  \eta^ {
\epsilon \zeta \}}I_{2w}(m^2)-\nonumber\\
&-16(5-w)(4-w)(3-w)(2-w) I^{\mu \nu \alpha \beta \gamma \delta \epsilon \zeta}_{2w}(m^2)\nonumber\\
& \equiv  \omega_{2w} \eta^{\{\mu \nu} \eta^{ \alpha \beta} \eta^ {\gamma \delta}\eta^ {\epsilon \zeta \}}.
\label{dif4}
\end{align}

In the expressions above, $2w$ is the degree of divergence of the integrals and  for the sake of brevity, we substitute the subscripts 
$log$, $quad$ and $quart$ by $0$, $2$ and $4$, respectively. Surface terms can be conveniently written as integrals of total 
derivatives, namely
\begin{eqnarray}
\upsilon_{2w}\eta^{\mu \nu}= \int_k\frac{\partial}{\partial k_{\nu}}\frac{k^{\mu}}{(k^2-m^2)^{2-w}},
\label{ts1}
\end{eqnarray}
\begin{eqnarray}
(\xi_{2w}-v_{2w})\eta^{\{ \mu \nu} \eta^{ \alpha \beta \}}= \int_k\frac{\partial}{\partial k_{\nu}}\frac{2(2-w)k^{\mu} k^{\alpha} k^{
\beta}}{(k^2-m^2)^{3-w}},
\label{ts2}
\end{eqnarray}
\begin{eqnarray}
&&(\sigma_{2w}-\xi_{2w})\eta^{\{ \mu \nu} \eta^{ \alpha \beta} \eta^ {\gamma \delta \}}=\nonumber \\&&
=\int_k\frac{\partial}{\partial k_{\nu}}\frac{4(3-w)(2-w)k^{\mu} k^{\alpha} k^{\beta} k^{\gamma} k^{\delta}}{(k^2-m^2)^{4-w}},
\label{ts3}
\end{eqnarray}
and
\begin{eqnarray}
&&(\omega_{2w}-\sigma_{2w})\eta^{\{ \mu \nu} \eta^{ \alpha \beta} \eta^ {\gamma \delta } \eta^ {\epsilon \zeta \}}=\nonumber \\=&&
\int_k\frac{\partial}{\partial k_{\nu}}\frac{8(4-w)(3-w)(2-w)k^{\mu} k^{\alpha} k^{\beta} k^{\gamma} k^{\delta}k^{\epsilon}k^{\zeta}}{(k^2-m^2)^{5-w}}.
\label{ts4}
\end{eqnarray}

We see that equations (\ref{dif1})-(\ref{dif4}) are undetermined because they are differences between divergent quantities. Each 
regularization scheme gives a different value for these terms. However, as physics should not depend on the schemes applied, we leave 
these terms to be arbitrary until the end of the calculation, fixing them by symmetry constraints or phenomenology, when it applies \cite
{Jackiw}.

It is noteworthy that this prescription is not the usual one since we do not evaluate divergent integrals or
regularization dependent quantities neither do we introduce a regulator or further parameters usually introduced in explicit 
regularization procedures. We do assume the existence of a regulator in order to give sense to the manipulation (\ref{2.1}). However, we 
do not say which one. That is because the introduction of an explicit regulator and additional parameters usually breaks symmetries
of the theory and it makes the renormalization procedure more laborious.

Besides of not modifying the theory, such as changing the dimension of space-time or breaking gauge or Lorentz symmetry spuriously, we 
can support or differ controversial results of the literature, which are most of the times caused by regularization dependent quantities
like surface terms in (\ref{dif1})-(\ref{dif4}). If those terms remain in the finite part of the amplitude, it can be arbitrary and
regularization dependent although of being finite. Therefore, we carry those terms till the end of the calculation and fix them using
a symmetry requirement, the fulfillment of a Ward identity, for instance.

We should also emphasize that although IR was consistently built for multi-loop calculation in scalar field theories \cite{Adriano}, its validity is questionable 
for arbitrary loop order in other theories and in curved space. If one works in a momentum space framework, one must ensure that causality and 
locality are guaranteed in all orders of perturbation theory. In differential regularization \cite{Freedman}, for instance, it was shown that those principles
hold at lower-order even in curved space \cite{Prange}. However, we do not worry with that in the present case since we perform only one-loop calculations.

\section{One-loop correction to the graviton propagator and the trace anomaly}
\label{3}

We consider the semi-classical approach of gravitation where matter fields are quantized in a classical curved background (see \cite
{Shapiro3} for a review). The action for scalar, fermion and Abelian vector are, respectively
\begin{eqnarray}
 S_s=\frac{1}{2}\int d^4 x \sqrt{-g}(g^{\mu \nu}\partial_{\mu}\phi\partial_{\nu}\phi+\xi R \phi^2), \label{3.1}\\
 S_f= i\int d^4 x e e^{\mu}_{a}\bar{\psi}\gamma^{a}D_{\mu}\psi
 \label{3.2}
\end{eqnarray}
and
\begin{eqnarray}
 S_v=-\frac{1}{4}\int d^4 x \sqrt{-g} F^{\mu \nu} F_{\mu \nu}.
 \label{3.3}
\end{eqnarray}
Where $e^{\mu}_{a}$ is the tetrad, $e=det\ e^{\mu}_{a}$ and $\xi$ is the non-minimal coupling.

In equation (\ref{3.2}), in order to couple fermions with the gravitational field, we need to define the covariant derivative,
\begin{equation}
D_{\mu} \psi= \partial_{\mu} \psi +\frac{1}{2}\omega_{\mu a b}\sigma^{ab} \psi,
\label{3.4}
\end{equation}
where $\omega_{\mu a b}$ is the spin connection, which depends on the tetrad, and $\sigma^{ab}=\frac{1}{4}[\gamma^a,\gamma^b]$, with $\gamma^a$ representing the Dirac matrices.

\begin{figure}
\centering
\begin{minipage}{.1\textwidth}
  \centering
  \includegraphics[trim=28mm 20mm 20mm 20mm, clip, scale=0.35,angle=-90]{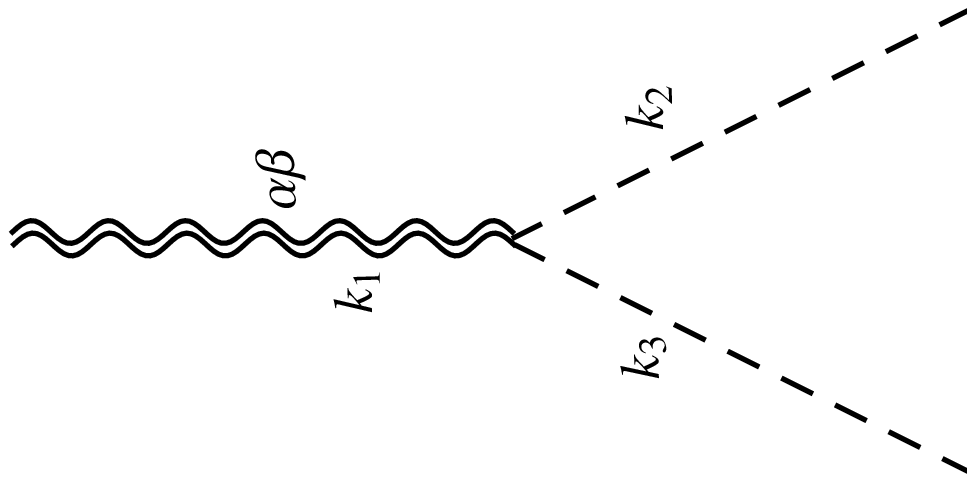}
\end{minipage}
\begin{minipage}{.9\textwidth}
  \centering
\end{minipage}
\begin{eqnarray*}
&&V_{s}^{\alpha \beta}(k_1,k_2,k_3)=\frac{i\kappa}{2}[\eta^{\alpha \beta}k_2\cdot k_3-k_2^{\alpha}k_3^{\beta}- k_3^{\alpha}k_2^{\beta} -\nonumber \\&& -2\xi(k_1^{\alpha}k_1^{\beta}-k_1^2\eta^{\alpha \beta})]
  \label{3.5}
\end{eqnarray*}	
\begin{minipage}{.1\textwidth}
  \centering
  \includegraphics[trim=28mm 20mm 20mm 20mm, clip, scale=0.35,angle=-90] {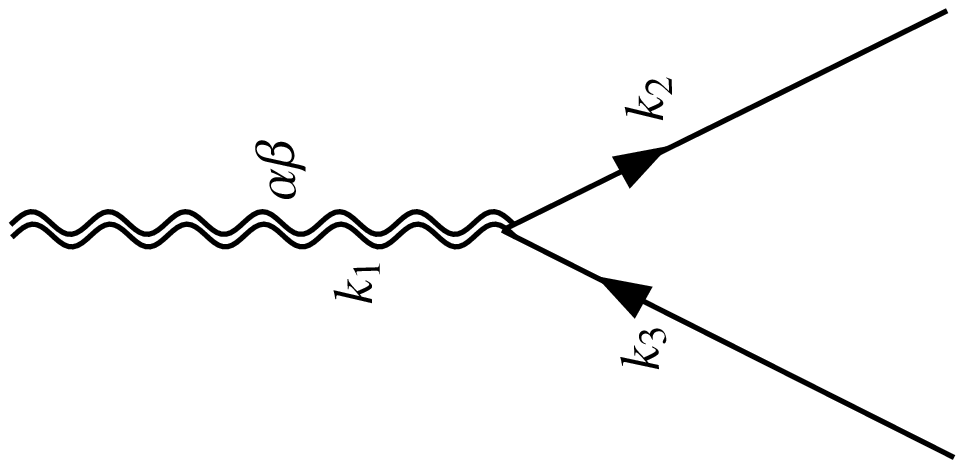}
\end{minipage}
\begin{minipage}{.9\textwidth}
  \centering
 \end{minipage}
\begin{eqnarray*}
&&V_{f}^{\alpha \beta}(k_2,k_3)=\frac{i\kappa}{8}[2\eta^{\alpha \beta}(\kslash_2+\kslash_3)-\nonumber \\&& - \gamma^{\alpha}(k_2+k_3)^{\beta}-\gamma^{\beta}(k_2+k_3)^{\alpha}] 
  \label{3.6}
	\end{eqnarray*}
\begin{minipage}{.1\textwidth}
  \centering
  \includegraphics[trim=27mm 20mm 20mm 20mm, clip, scale=0.35,angle=-90]{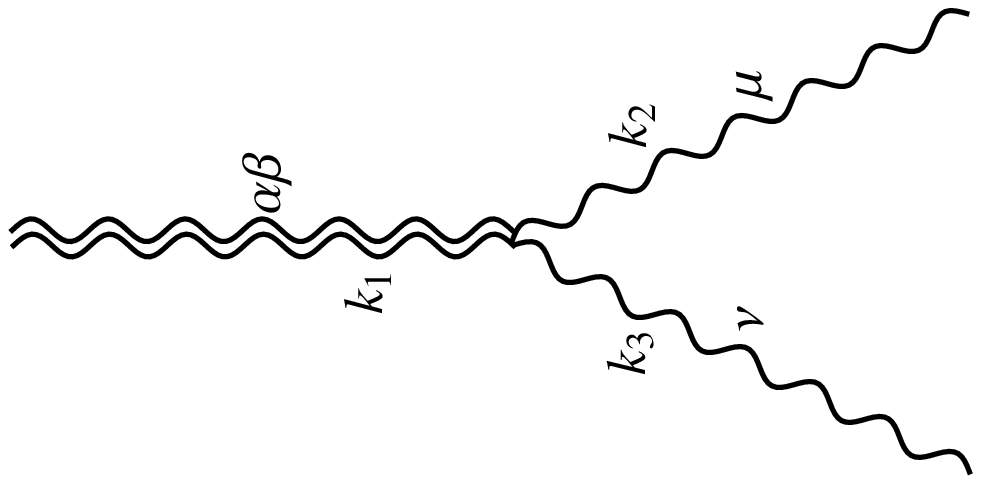}
\end{minipage}
\begin{minipage}{.9\textwidth}
  \centering
\end{minipage}
\begin{eqnarray*}
&&V_{v}^{\alpha \beta \mu \nu}(k_2,k_3)=\frac{i\kappa}{2} [(\eta^{\mu \{\alpha}\eta^{\beta\} \nu}-\eta^{\mu\nu}\eta^{\alpha\beta})k_2\cdot k_3 +(
\eta^{\alpha \beta
}k_2^{\nu}k_3^{\mu}+\nonumber \\&& +\eta^{\mu \nu}k_2^{\alpha}k_3^{\beta}
+\eta^{\mu \nu}k_2^{\beta}k_3^{\alpha}-\eta^{\alpha \nu}k_2^{\beta}k_3^{\mu}-\eta^{\mu \alpha}k_2^{\nu}k_3^{\beta}-\eta^{\mu \beta}k_2^{
\nu}k_3^{\alpha}-\nonumber \\&& -\eta^{\beta \nu}k_2^{
\alpha}k_3^{\mu})]
  \label{3.7}
	\end{eqnarray*}
\caption{Feynman rules for matter fields in linearized quantum gravity.}
\label{fig1}
\end{figure}

The actions expressed by (\ref{3.2}) and (\ref{3.3}) are classically conformal invariant and so is the action (\ref{3.1}) in the conformal limit 
$\xi\rightarrow1/6$. To compute the classical breaking we have first to calculate one-loop corrections to the graviton
propagator. In order to do this we first consider the weak field approximation, \textit{i.e.} we use the following expansions for the metric and 
the tetrad:
\begin{equation}
g^{\mu\nu}=\eta^{\mu\nu}+\kappa h^{\mu\nu}
\label{metricEXP}
\end{equation}
and
\begin{equation}
e_{\mu a}= \eta_{\mu a}+\frac{1}{2}{\kappa}h_{\mu a}.
\label{tetradEXP}
\end{equation}
Thus, using equations (\ref{metricEXP}) and (\ref{tetradEXP}) in (\ref{3.1}), (\ref{3.2}) and (\ref{3.3}), we obtain the Feynman rules up to first order in $\kappa$. We list them in Figure \ref{fig1}.

\begin{figure}
\centering
\includegraphics[trim=0mm 15mm 0mm 15mm,scale=0.9]{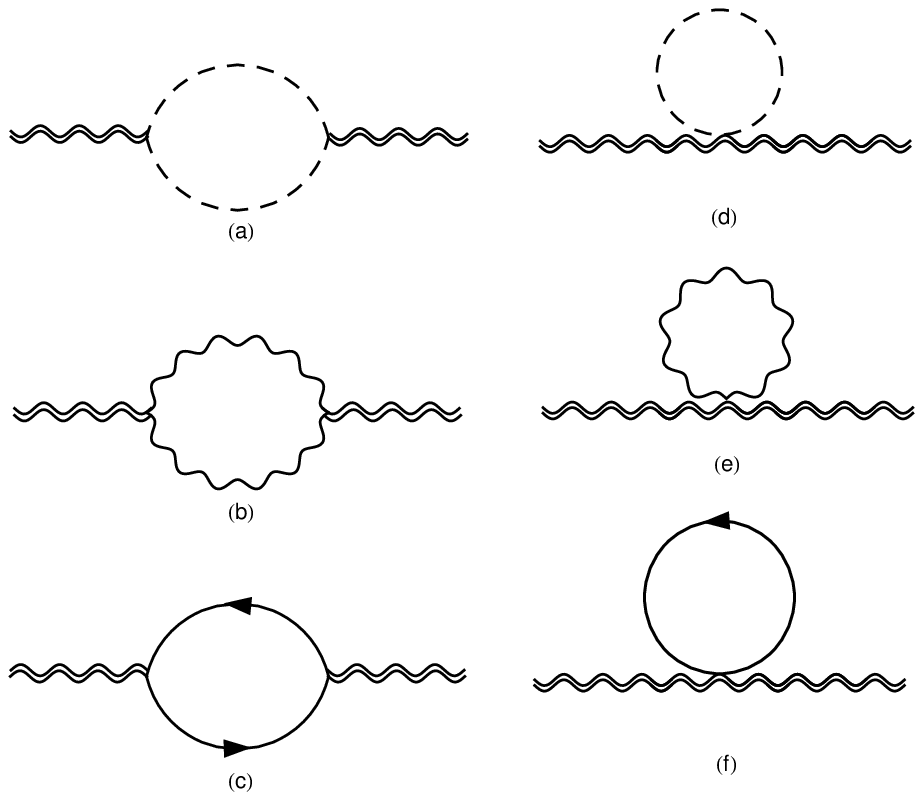}
\caption{One-loop corrections to the graviton propagator. The dashed, solid, waved and double-waved lines stand for scalar, fermion, 
vector and graviton, respectively.}
\label{fig2}
\end{figure}

Diagrams contributing to one-loop correction to the graviton propagator are presented in Figure \ref{fig2}. The finite part responsible
for the quantum breaking of conformal symmetry comes from diagrams $(a)$,$(b)$ and $(c)$. Loop diagrams $(d)$,$(e)$ and $(f)$ 
contribute only with quartic and quadratic divergences. Quadratic divergences for massless fields are made zero in dimensional
regularization \cite{Leibbrandt} and in Implicit Regularization \cite{Vieira}. Quartic divergences are unphysical in the sense that they 
do not contribute for physical quantities, like logarithmic divergences do when deriving the running of coupling constants, for 
instance. Both divergences also come from diagrams $(a)$,$(b)$ and $(c)$. Using symmetric integration, like 
$k^{\mu}k^{\nu}\rightarrow\frac{1}{4}\eta^{\mu\nu}k^2$, all quartic divergences can be transformed in a single form
$\int^{\Lambda}\frac{d^4k}{(2\pi)^4}$ which can be subtracted by a suitable cosmological counter-term (see \cite{DeWitt} for details).

Therefore, we have to calculate the following amplitudes:

\begin{eqnarray}
&&\Pi_{(a)}^{\mu \nu \alpha \beta}(p)=\frac{1}{2} \int_k V^{\mu \nu}_s(p,k,k+p)\frac{i}{k^2-m^2_s}\times \nonumber \\&& \times V^{\alpha \beta}_s(p,k+p,k)
\frac
{i}{(k+p)^2-m^2_s},
\label{ampli1}
\end{eqnarray}
\begin{eqnarray}
&&\Pi_{(b)}^{\mu \nu \alpha \beta}(p)=\frac{1}{2} \int_k  V^{\alpha \beta \lambda \theta}_v(k,k+p)\frac{-i\eta_{\lambda \gamma
}}{k^2-m^2_v}\times \nonumber \\&& \times V^{\mu \nu \gamma \delta}_v(k+p,k) \frac{-i\eta_{\delta \theta}}{(k+p)^2-m^2_v},
\label{ampli2}
\end{eqnarray}
\begin{eqnarray}
&&\Pi_{(c)}^{\mu \nu \alpha \beta}(p)= -\int_k Tr[ V^{\alpha \beta}_f(k,k+p)\frac{i}{\kslash-m_f}\times \nonumber \\&& \times V^{\mu \nu}_f(k+p,k)\frac{i}{\kslash+\pslash-m_f}].
\label{ampli3}
\end{eqnarray}

In the equations above, $1/2$ is a symmetry factor and we have introduced fictitious masses in the propagators. This is necessary because, although  the 
present integrals are infrared safe, expression (\ref{2.1}) without mass will break the original integral in two infrared divergent 
parts. The limit $m^2_i \to 0$ is taken in the end. In this process a renormalization scale $\lambda \ne 0$ appears. Observe that the other part of the massive vector propagator in equation (\ref{ampli2}) does not contribute since $k_{\lambda}k_{\gamma} V^{\alpha \beta 
\lambda \theta}_v(k,k+p)=0$ and $(k+p)_{\delta}(k+p)_{\theta}V^{\mu \nu \gamma \delta}_v(k+p,k)=0$. For sake of completeness, we list all regularized integrals coming from the expansion of equations (\ref{ampli1})-(\ref{ampli3}) in the appendix.

After taking the limits $m_s\rightarrow0$ and $\xi\rightarrow\frac{1}{6}$, we find that amplitude (\ref{ampli1}) is transverse up to surface 
terms defined in equations (\ref{dif1})-(\ref{dif4}):

\begin{widetext}
\begin{eqnarray}
&& \frac{2}{\kappa^2} p_{\alpha}\Pi_{(a)}^{\mu \nu \alpha \beta}(p)=\Bigg(\frac{37}{48}p^{\beta}\eta^{\mu\nu}p^2+\frac{1}{16}p^{\mu}\eta^{
\beta\nu
}p^2+\frac{1}{16}p^{\nu}\eta^{\beta\mu}p^2+\frac{2}{3}p^{\mu}p^{\beta}p^{\nu}\Bigg)p^2\upsilon_0-\Bigg(\frac{29}{48}p^{\beta}\eta^{\mu\nu
}p^2 +\frac{1}{16}p^{\mu}\eta^{\beta\nu}p^2+\nonumber \\&&+\frac
{1}{16}p^{\nu}\eta^{\beta\mu}p^2+\frac{1}{3}p^{\mu}p^{\beta}p^{\nu}\Bigg)p^2\xi_0 +\Bigg(\frac{109}{8}p^{\beta}\eta^{\mu\nu}p^2+\frac
{37}{8}p^{\mu}\eta^{\beta\nu}p^2+\frac{37}{8}p^{\nu}\eta^{\beta\mu}p^2 +\frac{121}{4}p^{\mu}p^{\beta}p^{\nu} \Bigg)p^2\sigma_0+\Bigg(\frac{1}{8}p^{\beta}\eta^{\mu\nu
}p^2+\nonumber \\&& -\frac{1}{8}p^{\mu}\eta^{\beta\nu}p^2+\frac{1}{8}p^{\nu}\eta^{\beta\mu}p^2+\frac{1}{2}p^{\mu}p^{\beta}p^{\nu}\Bigg)p^2\omega_0.
\label{3.8}
\end{eqnarray}
\end{widetext}

We see in equation (\ref{3.8}) that gauge invariance, {\it i. e.} $p_{\alpha}\Pi_{a}^{\mu \nu \alpha \beta}(p)=0$, holds if all 
indeterminacies expressed by surface terms are set to zero. That is  not the only possible solution. It may exist relations between 
these surface terms which also would make gauge invariance holds. In this case the finite result would be arbitrary \cite
{Jackiw,Scarpelli,Felipe,Gazzola}. We are going to discuss about this in section {\ref{s6}}. The ambiguity in the $a'$ term may be due 
to surface terms because they are often the source of ambiguities as we found out in other models \cite{Scarpelli,Felipe,Cabral}.

Our final result for the amplitude (\ref{ampli1}) is:
\begin{widetext}
\begin{eqnarray}
&&\frac{2}{\kappa^2}\Pi_{(a)}^{\mu \nu \alpha \beta}(p)=p^2(\eta^{\alpha\nu}\eta^{\beta\mu}p^2+\eta^{\alpha\mu}\eta^{\beta\nu}p^2 -p^{\alpha}p^{\mu}\eta^{\beta\nu} -p^{\beta}p^{\mu}\eta^{\alpha\nu}-p^{\alpha}p^{\nu}\eta^{\beta\mu}-p^{\beta}p^{\nu}\eta^{\alpha\mu})\Bigg[\frac{23}{1800}b +\frac{1}{240}\Bigg(I_{log}(\lambda^2)-\nonumber \\ && -b \ln \Bigg(-\frac{p^2}{\lambda^2}\Bigg)\Bigg)\Bigg]-p^2(\eta^{\alpha\beta}\eta^{\mu\nu}p^2-p^{
\alpha}p^{\beta}\eta^{\mu\nu}-p^{\mu}p^{\nu}\eta^{\alpha\beta})\Bigg[\frac{7}{675}b +\frac{1}{360}\Bigg(I_{log}(\lambda^2) -b \ln \Bigg(-\frac{p^2}{\lambda^2}\Bigg)\Bigg)\Bigg]+\frac{1}{180}\Bigg(\frac{41}{15}b+\nonumber \\&& +I_{log}(\lambda^2)-b \ln\Bigg(-\frac{p^2}{\lambda^2}\Bigg)\Bigg) p^{\alpha}p^{\beta}p^{\mu}p^{\nu}.
\label{3.9}
\end{eqnarray}
\end{widetext}

This result agrees with \cite{Shapiro4} after identifying $I_{log}({\lambda}^2)$ as the divergent part. We now return to the curvature 
tensor from weak field approximation. In order to do this we write the corresponding covariant expression and we focus on the $\Box R$ 
term
\begin{eqnarray}
&&S=\int d^4 x \sqrt{-g}(\alpha_1 C^2+\alpha_2 R^2 )\rightarrow \nonumber \\&& \rightarrow \int d^4 x \sqrt{-g}(2 \alpha_1 W+\alpha_2 R^2 ),
\label{action}
\end{eqnarray}
where we replace $C^2\rightarrow2W=2R^2_{\mu\nu}-\frac{2}{3}R^2$ since the Gauss-Bonnet topological invariant does
not contribute to the propagator since it reduces in a topological surface term.

Applying the definition of the energy-momentum tensor for the action (\ref{action}), we see that its trace is given by
\begin{equation}
\left\langle T^{\mu}_{\mu}\right\rangle=\frac{-2}{\sqrt{-g}}g^{\mu\nu}\frac{\delta S}{\delta g^{\mu\nu}}= 12 \alpha_2\Box R.
\label{traceT}
\end{equation}

Therefore, all we have to do is to determine the constant $\alpha_2$. For this purpose, we write $W$ and $R^2$ in the weak field limit
up to second order in $\kappa$:
\begin{widetext}
\begin{eqnarray}
&&\int d^4 x \sqrt{-g}R^2= \int d^4 x h^{\mu\nu}[\partial_{\mu}\partial_{\nu}\partial_{\alpha}\partial_{\beta}  +\eta_{
\mu\nu}\eta_{\alpha\beta}\partial^4-(\eta_{\mu\nu}\partial_{\alpha}\partial_{\beta}\partial^2+\eta_{\alpha\beta}\partial_{\mu}\partial_{\nu}
\partial^2)]h^{\alpha\beta}
\label{3.10}
\end{eqnarray}
\begin{eqnarray}
&&\int d^4 x \sqrt{-g}W= \int d^4 x h^{\mu\nu}\Bigg[\frac{1}{6}\partial_{\mu}\partial_{\nu}\partial_{\alpha}\partial_{\beta}-\frac{1}{12}\eta_{\mu\nu}\eta_{\alpha\beta}\partial^4+\frac{1}{12}(\eta_{\mu\nu}\partial_{\alpha}\partial_{\beta}\partial^2+\eta_{\alpha\beta}\partial_{\mu}
\partial_{\nu}\partial^2)
 +\frac{1}{8}(\eta_{\mu\alpha}\eta_{\nu\beta}+\eta_{\nu\alpha}\eta_{\mu\beta})\partial^4-\nonumber \\&& -\frac{1}{8}\eta_{\mu\alpha}\partial_{\nu}
\partial_{\beta}\partial^2 \Bigg]h^{\alpha\beta}.
\label{3.10.1}
\end{eqnarray}
\end{widetext}

Replacing eqs. (\ref{3.10}) and (\ref{3.10.1}) in eq. (\ref{action}) and comparing with eq. (\ref{3.9}) written in the 
position space (the action for the graviton propagator is $S=-\frac{1}{2}\int d^4 x h^{\mu \nu}\bar{\Pi}_{\mu\nu\alpha\beta}h^{\alpha 
\beta}$, where $\bar{\Pi}_{\mu\nu\alpha\beta}$ is the Fourier transform of eq. (\ref{3.9})), we get 
$12\alpha_2=\frac{1}{180(4\pi)^2}$. Therefore, our result for the anomaly is
\begin{equation}
\left\langle T^{\mu}_{\mu}\right\rangle_{scalar}=\frac{1}{180(4\pi)^2}\Box R
\end{equation}

This result agrees with the one obtained in \cite{Duff3,Hawking,Christensen,Shapiro4}, where it was applied dimensional 
regularization, $\zeta-$function regularization, point-splitting regularization and proper time cut-off regularization, respectively. We 
proceed using the same idea to obtain the anomaly contributions coming from vector and spinor fields. The result of the one-loop 
correction to the graviton propagator for the amplitudes (\ref{ampli2}) and (\ref{ampli3}) are, respectively (if we set again the surface terms to zero, gauge invariance holds, \textit{i.e.} $p_{\alpha}\Pi_{(b)}^{\mu \nu \alpha \beta}(p)=0$ and $p_{\alpha}\Pi_{(c)}^{\mu \nu \alpha \beta}(p)=0$)
\begin{widetext}
\begin{align}
&&\frac{2}{\kappa^2}\Pi_{(b)}^{\mu \nu \alpha \beta}(p)=p^2(\eta^{\alpha\nu}\eta^{\beta\mu}p^2+\eta^{\alpha\mu}\eta^{\beta\nu}p^2-p^{\alpha}p^{\mu}
\eta^{\beta\nu}-p^{\beta}p^{\mu}\eta^{\alpha\nu}-p^{\alpha}p^{\nu}\eta^{\beta\mu}-p^{\beta}p^{\nu}\eta^{\alpha\mu})\Bigg[\frac{4}{75}b
 +\frac{1}{20}\Bigg(I_{log}(\lambda^2)-\nonumber \\&& -b \ln \Bigg(-\frac{p^2}{\lambda^2}\Bigg)\Bigg)\Bigg]-p^2(\eta^{\alpha\beta}\eta^{\mu\nu}p^2 -p^{
\alpha}p^{\beta}\eta^{\mu\nu}-p^{\mu}p^{\nu}\eta^{\alpha\beta})\Bigg[\frac{1}{450}b+\frac{1}{30}\Bigg(I_{log}(\lambda^2) -b \ln \Bigg(-\frac{p^2}{\lambda^2}\Bigg)\Bigg)\Bigg]+\frac{1}{15}\Bigg(\frac{47}{30}b+\nonumber \\&& +I_{log}(
\lambda^2)-b \ln \Bigg(-\frac{p^2}{\lambda^2}\Bigg)\Bigg) p^{\alpha}p^{\beta}p^{\mu}p^{\nu}
\label{3.11}
\end{align}
\end{widetext}
and
\begin{widetext}
\begin{align}
&\frac{2}{\kappa^2}\Pi_{(c)}^{\mu \nu \alpha \beta}(p)=p^2(\eta^{\alpha\nu}\eta^{\beta\mu}p^2+\eta^{\alpha\mu}\eta^{\beta\nu}p^2-p^{\alpha}p^{\mu}\eta^{\beta\nu} -p^{\beta}p^{\mu}\eta^
{\alpha\nu}-p^{\alpha}p^{\nu}\eta^{\beta\mu}-p^{\beta}p^{\nu}\eta^{\alpha\mu})\Bigg[
\frac{3}{50}b +\frac{1}{40}\Bigg(I_{log}(\lambda^2)-\nonumber \\&-b \ln \Bigg(-\frac{p^2}{\lambda^2}\Bigg)\Bigg)\Bigg]-p^2(\eta^{\alpha\beta}\eta^{\mu\nu}p^2 -p^{
\alpha}p^{\beta}\eta^{\mu\nu}-p^{\mu}p^{\nu}\eta^{\alpha\beta})\Bigg[\frac{23}{450}b+\frac{1}{60}\Bigg(I_{log}(\lambda^2) -b \ln \Bigg(-\frac{p^2}{\lambda^2}\Bigg)\Bigg)\Bigg]+\frac
{1}{30}\Bigg(\frac{31}{15}b+ \nonumber \\& +I_{log}(\lambda^2)-b \ln \Bigg(-\frac{p^2}{\lambda^2}\Bigg)\Bigg) p^{\alpha}p^{\beta}p^{\mu}p^{\nu}.
\label{3.12}
\end{align}
\end{widetext}

The corresponding values for the constant $\alpha_2$ for equations (\ref{3.11}) and (\ref{3.12}) are $12\alpha_2=-\frac{1}{10(4\pi)^2}$ and 
$12\alpha_2=\frac{1}{30(4\pi)^2}$, respectively. Multiplying each diagram by the number of fields, our final result is
\begin{eqnarray}
 \langle T^{\mu}_{\mu} \rangle=&&\left(\frac{1}{180(4\pi)^2}N_s+\frac{1}{30(4\pi)^2}N_f-\frac{1}{10(4\pi)^2}N_v\right)\Box R=\nonumber \\&& =\beta_3 \Box R. 
\label{resultado}
\end{eqnarray}

Therefore, we found out that apparently no ambiguity appears in the massless case and if we require gauge symmetry that fixes the 
surface terms to zero. This result agrees with all regularization methods \cite{Hawking,Christensen,Vilkovisky2} but the one 
obtained by dimensional regularization \cite{Shapiro2,Duff3}. This may suggest the result (\ref{resultado}) is universal and
dimensional regularization provides a different result because of a hard breaking of conformal symmetry. However, as we are going
to see in section \ref{4}, there is an inherent ambiguity associated with the renormalization. Moreover, in section \ref{s6} we 
present how this anomaly can be plagued by the arbitrary surface term, which also makes its result ambiguous.

\section{Renormalization}
\label{4}

We now perform the one-loop renormalization. Therefore, we write the one-loop renormalized action corresponding to the calculation of the
previous section
\begin{equation}
S_R(a_1,a_2)=S_{vacuum}(a^{(0)}_1,a^{(0)}_2)+\bar{\Gamma}^{(1)}+\Delta S_{vacuum},
\label{4.1}
\end{equation} 
where $S_{vacuum}(a^{(0)}_1,a^{(0)}_2)=\int d^4 x \sqrt{-g}(a^{(0)}_1 C^2+a^{(0)}_2 R^2)$ is the vacuum action, $\bar{\Gamma}^{(1)}$ is 
the one-loop effective action and $\Delta S_{vacuum}$ is the counter-term action.

In order to renormalize, we seize the results of section \ref{3}. Considering, for instance, the photon correction given by equation (\ref{3.11}), we have the following effective action
\begin{eqnarray}
&&\bar{\Gamma}^{(1)}=\frac{1}{(4\pi)^2}\int d^4 x \sqrt{-g}\Big[ C_{\mu\nu\alpha\beta}\left(-\frac{4}{75} +\right. \nonumber \\ && \left. +\frac{1}{20}
\left((4\pi)^2 i I_{log}(\lambda^2)+\ln\left(\frac{\Box}{\lambda^2}\right)\right)\right)C^{\mu\nu\alpha\beta}-\nonumber \\&&-\frac{1}{120}R^2\Big].
\label{4.2}
\end{eqnarray} 

We may choose $\Delta S_{vacuum}=-\frac{i}{20}\int d^4 x \sqrt{-g}I_{log}(\lambda^2)C^2$. We add this counter-term in order to 
remove the divergent integral $I_{log}(\lambda^2)$. This is equivalent to the {\it MS} renormalization scheme as we
have shown in reference \cite{Marcos}. However, it is also possible to add a {\it finite} local counter-term of the form
$\frac{1}{(4\pi)^2}\alpha \int d^4 x \sqrt{-g} R^2$ since it is a vacuum term  and it does not break conformal symmetry of the quantum fields. Considering these 
counter-terms, we end up with the following renormalized action
\begin{eqnarray}
&& S_R(a_1,a_2)=\frac{1}{(4\pi)^2}\int d^4 x \sqrt{-g}\Big[ C_{\mu\nu\alpha\beta}\Big( a^{(0)}_1-\frac{4}{75} +\nonumber \\&& +\frac{1}{20}\ln\left(\frac{\Box}{\lambda^2}\right) \Big)C^{\mu\nu\alpha\beta}+\Big(a^{(0)}_2
  -\frac{1}{120}+\alpha \Big)R^2\Big]=\nonumber \\&& =\int d^4 x \sqrt{-g}(a_1 C^2+a_2 R^2).
\label{4.1}
\end{eqnarray} 

Requiring that equation (\ref{4.1}) must not depend on the renormalization group scale $\lambda$, we find the one-loop $\beta$-function
\begin{equation}
\beta_1=\lambda \frac{\partial a_1}{\partial \lambda}=2\lambda^2 \frac{\partial a_1}{\partial \lambda^2}=-\frac{1}{10(4\pi)^2}.
\label{4.2}
\end{equation}

Following the same idea, the contributions coming from the scalar and the spinor field are $\beta_1=-\frac{1}{120(4\pi)^2}$ and 
$\beta_1=-\frac{1}{20(4\pi)^2}$, respectively. This result agrees with \cite{Shapiro4,Shapiro6} where it was applied the $\overline{MS}$ 
scheme. In this case we have found only the ultraviolet behavior of the $\beta$-function since we consider massless matter fields in a
curved background.

Clearly, the addition of the local finite counter-term generates an arbitrariness in the conformal anomaly. Applying equation (\ref
{traceT}) for the action (\ref{4.1}) we find the arbitrary result 
\begin{equation}
\left\langle T^{\mu}_{\mu}\right\rangle_{vector}=\frac{1}{(4\pi)^2}\left(-\frac{1}{10}+12\alpha\right)\Box R
\label{4.3}
\end{equation}

The result of equation (\ref{4.3}) is compatible with regularization schemes which breaks hardly conformal symmetry such as dimensional
regularization, as mentioned before. The result also agrees with the obtained in Pauli-Villars regularization, where an ambiguous 
result can also be found for the massive theory \cite{Shapiro5,Shapiro6}. In the next section, we show that an arbitrariness also appears
if we do not set all surfaces terms to zero.

\section{Arbitrariness in the conformal anomaly}
\label{s6}

We return to the previous amplitudes in order to see what happens if we do not set all surfaces terms to zero.
For instance, consider again the amplitude (\ref{ampli2}). However, this time we investigate if there is a relation
between surface terms which also make the final amplitude gauge invariant. As before we use the gauge Ward identity in order to fix 
those arbitrary surface terms. After taking the limit $m\rightarrow0$, we find that amplitude (\ref{ampli2}) is transverse 
up to surfaces terms

\begin{widetext}
\begin{eqnarray}
&& \frac{2}{\kappa^2} p_{\alpha}\Pi_{(b)}^{\mu \nu \alpha \beta}(p)=\Bigg(\frac{1}{8}p^{\beta}\eta^{\mu\nu}p^2+p^{\mu}\eta^{
\beta\nu
}p^2+p^{\nu}\eta^{\beta\mu}p^2+p^{\mu}p^{\beta}p^{\nu}\Bigg)p^2\upsilon_0-\Bigg(\frac{1}{8}p^{\beta}\eta^{\mu\nu
}p^2 +\frac{3}{4}p^{\mu}\eta^{\beta\nu}p^2+\nonumber \\&&+\frac
{3}{4}p^{\nu}\eta^{\beta\mu}p^2+\frac{1}{2}p^{\mu}p^{\beta}p^{\nu}\Bigg)p^2\xi_0 +\Bigg(\frac{37}{4}p^{\beta}\eta^{\mu\nu}p^2+\frac
{73}{4}p^{\mu}\eta^{\beta\nu}p^2+\frac{73}{4}p^{\nu}\eta^{\beta\mu}p^2 +\frac{121}{2}p^{\mu}p^{\beta}p^{\nu} \Bigg)p^2\sigma_0-\Bigg(\frac{1}{4}p^{\beta}\eta^{\mu\nu
}p^2+\nonumber \\&& +\frac{1}{4}p^{\mu}\eta^{\beta\nu}p^2+\frac{1}{4}p^{\nu}\eta^{\beta\mu}p^2+p^{\mu}p^{\beta}p^{\nu}\Bigg)p^2\omega_0=0
\label{6.1}
\end{eqnarray}
\end{widetext}

As in equation (\ref{3.8}), setting all surface terms to zero is a possible solution. However, we can easily see that it is possible to 
establish a relation between them which would also satisfy (\ref{6.1}). Considering the tensorial structure, we see that requiring gauge 
invariance gives us the relations
\begin{eqnarray}
\upsilon_0-\xi_0+74\sigma_0-2\omega_0=0,\\
4\upsilon_0-3\xi_0+73\sigma_0-\omega_0=0,\\
2\upsilon_0-\xi_0+121\sigma_0-2\omega_0=0. \nonumber\\
\label{6.2}
\end{eqnarray}

Since the parameters are overdetermined by equations above we may write:
$\upsilon_0=-47\sigma_0$, $\xi_0=-\frac{257}{5}\sigma_0$ and $\omega_0=\frac{196}{5}\sigma_0$. That means that gauge invariance was not 
sufficient to fix all the arbitrary terms. Consequently, we can replace $\upsilon_0$, $\xi_0$ and $\omega_0$ in the amplitude and the 
final answer now depends on the arbitrary surface term $\sigma_0$. As a result, the anomaly become arbitrary because it depends on the 
arbitrary surface term
\begin{align}
\left\langle T^{\mu}_{\mu}\right\rangle_{vector}=-\frac{1}{(4\pi)^2}\left(\frac{1}{10}+\frac{497}{15}\sigma_0\right)\Box R=\nonumber\\
=\frac{1}{(4\pi)^2}\left(-\frac{1}{10}+\sigma'_0\right)\Box R
\label{6.3}
\end{align}

This result is compatible with the arbitrariness that appears in renormalization, as presented in the previous section. It also
agrees with the result found in dimensional regularization of Ref. \cite{Shapiro2} and  in Pauli-Villars regularization \cite{Shapiro2,Shapiro5,Shapiro6}.

In order to support our result, we also calculated the anomaly for the massive case. In this case, we found the same ambiguity that 
appeared in (\ref{6.3}) (massless case), according to \cite{Shapiro2,Shapiro5,Shapiro6}. Although in the latter, the ambiguity was found 
only in the massive theory, our result shows that the ambiguity in the conformal anomaly appears even in the massless case.
\\

\section{Conclusion}
\label{5}

In this paper, we considered an implicit momentum space regularization derivation of the one-loop conformal anomaly in order to shed 
some light on controversies raised in the literature in which some finite breaking terms are ambiguous. Our approach is 
specially tailored to study quantum symmetry breakings. In this approach, regularization dependent indeterminacies expressed by surface 
terms are identified to be fixed on symmetry grounds. However, as in the present case the symmetry content of the theory is 
not sufficient to fix all the arbitrary terms and the finite part of the amplitude is ambiguous and regularization dependent although of 
being finite. As a result, we find out that there is an unavoidable arbitrariness in the conformal anomaly even in the massless case. 
Our result is equivalent to the usual subtraction procedure of including an $\int d^4 x \sqrt{-g} R^2$-term in the renormalized action.
\\

{\bf Acknowledgements}

The authors are grateful to Ilya Shapiro for clarifying discussions. A. R. V. acknowledges financial support by CNPq. M. S. acknowledges research grants from CNPq and Durham University 
for the kind hospitality. J. C. C. F. acknowledges financial support by CAPES. G. G. acknowledges financial support by FAPEMIG. 
This work is dedicated to the memory of professor M. C. Nemes.

\begin{widetext}

\section*{Appendix} \label{A}
The results of the regularized integrals in the massless limit are:

\begin{align}
&\int_k \frac{1}{k^2(k+p)^2}= I_{log}(\lambda^2)+2 b -b \ln \left(-\frac{p^2}{\lambda^2}\right),\\
&\int_k \frac{k^2}{k^2(k+p)^2}= -p^2 \upsilon_0,\\
&\int_k \frac{k^2 k^{\alpha}}{k^2(k+p)^2}= p^2 p^{\alpha}(\xi_0-\upsilon_0),\\
&\int_k \frac{k^4}{k^2(k+p)^2}= p^4 (3\upsilon_0-2\xi_0),\\
&\int_k \frac{k^{\alpha}}{k^2(k+p)^2}= \frac{1}{2}p^{\alpha}\left[-I_{log}(\lambda^2)+\upsilon_0-2 b +b \ln \left(-\frac{p^2}{\lambda^2}\right)\right] , \\
&\int_k \frac{k^{\alpha}k^{\beta}}{k^2(k+p)^2}= \left(\frac{1}{3} p^{\alpha}p^{\beta}-\frac{1}{12} p^2 \eta^{\alpha\beta}\right)\left
[I_{log}(\lambda^2)-b\ln \left(-\frac{p^2}{\lambda^2}\right)\right]-\left(\frac{1}{3} p^{\alpha}p^{\beta}+\frac{1}{6}p^2 \eta^{\alpha\beta}\right)\xi_0 + \nonumber\\
&+\frac{1}{4} p^2\eta^{\alpha\beta} \upsilon_0 + \frac{13}{18}b p^{\alpha}p^{\beta}-\frac{2}{9}p^2 b \eta^{\alpha\beta} ,\\
&\int_k \frac{k^{\mu}k^{\alpha}k^{\beta}}{k^2(k+p)^2}= \frac{1}{24}(p^{\{ \mu}\eta^{\alpha \beta\}}p^2-6 p^{\alpha}p^{\beta}p^{\mu})\left[I_{log}(\lambda^2)-\xi_0-b \ln \left(-\frac{p^2}{\lambda^2}\right)\right]+\nonumber\\ 
&+3(p^{\{ \mu}\eta^{\alpha \beta\}}p^2+2 p^{\alpha}p^{\beta}p^{\mu})\sigma_0+\frac{1}{9}b p^{\{ \mu}\eta^{\alpha \beta\}}p^2-\frac{7}{12}b p^{\mu}p^{\alpha}p^{\beta},\\
&\int_k \frac{k^2 k^{\alpha}k^{\beta}}{k^2(k+p)^2}=\frac{1}{4}p^4\eta^{\alpha\beta}(\xi_0-\upsilon_0)-6 p^2(4p^{\alpha}p^{\alpha}+\eta^{\alpha\beta}p^2)\sigma_0,\\
&\int_k \frac{k^{\mu}k^{\nu} k^{\alpha}k^{\beta}}{k^2(k+p)^2}= \frac{1}{240}(\eta^{\{\mu \nu}\eta^{\alpha \beta \}}p^4-3 p^2 p^{\{ \mu}p^{\nu}\eta^{\alpha \beta\}}+48p^{\alpha}p^{\beta}p^{\mu}p^{\nu} )\left[I_{log}(\lambda^2)-b \ln \left(-\frac{p^2}{\lambda^2}\right)\right]\nonumber\\
&+\frac{1}{48}\eta^{\{\mu \nu}\eta^{\alpha \beta \}}p^4\left(26\sigma_0-\xi_0-\frac{6}{5}\omega_0\right)+\frac{1}{48}p^{\{\alpha}p^{\beta}\eta^{\mu \nu \}}p^2\left(26\sigma_0+\xi_0-\frac{12}{5}\omega_0\right)\nonumber\\
&+\frac{1}{600}b\left(\frac{23}{3}p^4\eta^{\{\alpha \beta}\eta^{\mu \nu \}}-\frac{41}{2} p^2 p^{\{\alpha}p^{\beta}\eta^{\mu \nu \}}\right)+\frac{149}{300}b p^{\mu}p^{\nu}p^{\alpha}p^{\beta}-\frac{1}{5}p^{\mu}p^{\nu}p^{\alpha}p^{\beta}\omega_0,\\
\end{align}
where $\lambda$ is the renormalization group scale and $b \equiv \frac{i}{(4\pi)^2}$. For the sake of simplicity we omit quartic
divergent integrals. The surface terms are defined in Section \ref{2}.

\end{widetext}


\begin{thebibliography}{99}
\bibitem{Livro} I. L. Buchbinder, S. D. Odintsov and I. L. Shapiro, {\it Effective Action in Quantum Gravity}, Bristol: Institute of Physics Publishing (1992).
\bibitem{Maldacena} J. Maldacena, Int. J. Theor. Phys. {\bf 38}, 1113 (1999).
\bibitem{Makoto} M. Natsuume, hep-th/14093575
\bibitem{ABJ} J. S.  Bell and R. Jackiw, 1969 Nuovo Cimento. {\bf 60}, 47 (1969); S. L. Adler, Phys. Rev. {\bf 177} 2426 (1969).
\bibitem{Duff1} D. M. Capper, M. J. Duff and L. Halpern, Phys. Rev.  D {\bf 10}, 461 (1974).
\bibitem{Brown} L. S. Brown, Phys. Rev. D {\bf 15}, 1469 (1977) .
\bibitem{Jackiw} R. Jackiw, Int. J. Mod. Phys. B {\bf 14}, 2011 (2000).
\bibitem{IRA} L. C. Ferreira, A. L. Cherchiglia, B. Hiller, M. Sampaio and M. C. Nemes, Phys. Rev. D {\bf 86}, 025016 (2012) .
\bibitem{tHooft} G. 't Hooft and M. Veltman, Nucl. Phys. B {\bf 44}, 189 (1972).
\bibitem{IR} O. A. Battistel, A. L. Mota and M. C. Nemes, Mod. Phys. Lett. A {
\bf 13}, 1597 (1998).
\bibitem{Leo} L. A. M. Souza, M. Sampaio and M. C. Nemes, Phys. Lett. B {\bf 632}, 717 (2006).
\bibitem{review} M. J. Duff, Class. Quantum Grav. {\bf 11}, 1387 (1994).
\bibitem{Shapiro2} M. Asorey, E. V. Gorbar and I. L. Shapiro, { Class. Quantum Grav.} {\bf 21}, 163 (2004).
\bibitem{Duff2} D. M. Capper and M. J. Duff, Nucl. Phys. B {\bf 82}, 147 (1974).
\bibitem{Duff3} D. M. Capper and M. J. Duff, Nuovo Cimento A {\bf 23}, 173 (1974).
\bibitem{Duff4} D. M. Capper and M. J. Duff, Phys. Lett. A {\bf 53}, 361 (1975).
\bibitem{Kallosh} R. Kallosh, Phys. Lett. B {\bf 55}, 321 (1974).
\bibitem{Fradkin}  E. S. Fradkin and G. A. Vilkovisky, Phys. Lett. B {\bf 73}, 209 (1978).
\bibitem{Adler} S. L. Adler, J. Lieberman and Y. J. Ng, { Ann. Phys.} {\bf 106}, 279 (1977).
\bibitem{Englert} F. Englert, C. Truffin, R. Gastmans, Nucl. Phys. B {\bf 117}, 407 (1976).
\bibitem{Antoniadis1} I. Antoniadis and N. C. Tsamis,  Phys. Lett. B {\bf 144}, 55 (1984).
\bibitem{Antoniadis2} I. Antoniadis, J. Iliopoulus and T. N. Tomaras, Nucl. Phys. B {\bf 261} 157 (1985).
\bibitem{Antoniadis3} I. Antoniadis, C. Kounnas and  D. V. Nanopoulos, Phys. Lett. B {\bf 162}, 309 (1985).
\bibitem{Bollini} C. G. Bollini and J. J. Giambiagi, Nuovo Cimento B {\bf 12}, 20 (1972).
\bibitem{Hawking} S. Hawking, Commun. Math. Phys {\bf 55}, 133 (1977).
\bibitem{Christensen} S. M.  Christensen, Phys. Rev. D {\bf 14}, 2490 (1976); S. M. Christensen, Phys. Rev. D {\bf 17}, 946 (1978).
\bibitem{Vilkovisky2} A. O. Barvinsky and G. A. Vilkovisky, Nucl. Phys. B {\bf 333}, 471 (1990);  A. O. Barvinsky, Y. V. Gusev, G. A. Vilkovisky and V. V. 
Zhitnikov, Nucl. Phys. B {\bf 439}, 561 (1995).
\bibitem{Henningson} M. Henningson and K.  Skenderis, { J. High Energy Phys.} JHEP07 (1998) 023.
\bibitem{Nicolas} N. Boulanger, { J. High Energy Phys.} JHEP07 (2007) 069;  N. Boulanger, { Phys. Rev. Lett.} {\bf 98}, 261302 (2007). 
\bibitem{Christensen2} S. M. Christensen and S. A. Fulling, Phys. Rev. D {\bf 15}, 2088 (1977).
\bibitem{Balsan} R. Balbinot, A. Fabbri and I. Shapiro, Nucl. Phys. B {\bf 559}, 301 (1999); R. Balbinot, A. Fabbri and I.L. Shapiro, {Phys. Rev. Lett.} {\bf 
83} 1494 (1999).
\bibitem{Starobinsky} A. A. Starobinsky, Phys. Lett. B {\bf 91}, 99 (1980).
\bibitem{Shapiro} I. L. Shapiro, Int. J. Mod. Phys {\bf 11}, 1159 (2002).
\bibitem{Sergei} S. Nojiri and S. D. Odintsov, Phys. Rev. D {\bf 59}, 044026 (1999). 
\bibitem{Sergei2} I. H. Brevik and S. D. Odintsov, Phys. Lett. B {\bf 475}, 247 (2000).
\bibitem{Sergei3} S. Nojiri and S. D. Odintsov, Phys. Lett. B {\bf 484}, 119 (2000). 
\bibitem{Shapiro5} M. Asorey, G. Berredo-Peixoto and  I. L. Shapiro, Phys. Rev. D {\bf 74}, 124011 (2006).
\bibitem{Shapiro3} I. L.  Shapiro, { Class. Quantum Grav.} {\bf 25}, 103001 (2008).
\bibitem{Adriano} A. L. Cherchiglia, M. Sampaio and  M. C. Nemes, Int. J. of Mod. Phys. A {\bf 26}, 2591 (2011).
\bibitem{Freedman} D. Z. Freedman, K. Johnson and J. I. Latorre, Nucl. Phys. B {\bf 371}, 353 (1999).
\bibitem{Prange} D. Prange, { J. Phys. A: Math. Gen.} {\bf 32}, 2225 (1999).
\bibitem{Leibbrandt} G. Leibbrandt, Rev. Mod. Phys. {\bf 47}, 849 (1975).
\bibitem{Vieira} A. L. Cherchiglia, A. R.  Vieira,  B. Hiller, M. Sampaio and A. P. Ba\^eta Scarpelli, { Ann. Phys.} {\bf 351}, 751 (2014).
\bibitem{DeWitt} R. Utiyama and B. S. DeWitt, J. of Math. Phys. {\bf 3}, 608 (1962).
\bibitem{Scarpelli} A. P. Ba\^eta Scarpelli, M. Sampaio, M. C. Nemes and B. Hiller, { Eur. Phys. J. C}  {\bf 56}, 571 (2008).
\bibitem{Felipe} J. C. C. Felipe, A. R. Vieira, A. L. Cherchiglia, A. P. Ba\^eta Scarpelli and M. Sampaio, Phys. Rev.  D {\bf 89}, 105034 (2014).
\bibitem{Gazzola} G.  Gazzola, H. G. Fargnoli, A. P. Ba\^eta Scarpelli, M. Sampaio, M. C. Nemes, { J. Physics G} {\bf 39}, 035002 (2012).
\bibitem{Cabral} A. L. Cherchiglia, L. A. Cabral, M. C. Nemes and M. Sampaio, Phys. Rev. D {\bf 87}, 065011 (2013).
\bibitem{Marcos} M. Sampaio, A. P. Ba\^eta Scarpelli, B. Hiller, A. Brizola, M. C. Nemes and S. Gobira, Phys. Rev. D {\bf 65}, 125023 (2002).
\bibitem{Shapiro4} E. V. Gorbar and I. L. Shapiro, J. High Energy Phys. 02 (2003) 021.
\bibitem{Shapiro6} E. V. Gorbar and I. L.  Shapiro, J. High Energy Phys. 06 (2003) 004.
\end{thebibliography}
\end{document}